# Creating Usable Pin Array Tactons for Non-Visual Information

Thomas Pietrzak‡, Andrew Crossan†, Stephen A. Brewster†, Benoît Martin‡ and Isabelle Pecci‡

**Abstract**— Spatial information can be difficult to present to a visually impaired computer user. In this paper we examine a new kind of tactile cueing for non-visual interaction as a potential solution, building on earlier work on vibrotactile Tactons. However, unlike vibrotactile Tactons, we use a pin array to stimulate the finger tip. Here, we describe how to design static and dynamic Tactons by defining their basic components. We then present user tests examining how easy it is to distinguish between different forms of pin array Tactons demonstrating accurate Tacton sets to represent directions. These experiments demonstrate usable patterns for static, wave and blinking pin array Tacton sets for guiding a user in one of eight directions. A study is then described that shows the benefits of structuring Tactons to convey information through multiple parameters of the signal. By using multiple independent parameters for a Tacton, this study demonstrates participants perceive more information through a single Tacton. Two applications using these Tactons are then presented: a maze exploration application and an electric circuit exploration application designed for use by and tested with visually impaired users.

**Index Terms**—D.2.14.a User interfaces, H.1.2.a Human factors, H.5.2.G. Haptic I/O, K.3.1.A Computer-assisted instruction

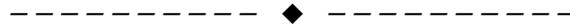

## 1 INTRODUCTION

PRESENTING information that is visually oriented to blind and visually impaired people is a challenging, but important problem. For many educational applications or for browsing information on the Internet, it is common to use representations that mainly use the visual channel. For example maps, charts, or tables present information through the relative positions of objects. It is important for a visually impaired person to be able to explore data where the spatial component of the information is key to its understanding. To allow a visually impaired user to browse spatial information, techniques must be examined to guide them through the data.

Here we consider how technology can impact on this problem. Computers are now starting to play a far greater role in education in schools. However, computer users rely heavily on visual feedback, with the graphical user interface playing an important part in interactions. Users with little or no vision must rely on other modalities to access the same information. Screen readers, such as JAWS from Freedom Scientific (freedomscientific.com), have proved to be a successful solution for accessing the textual information required to interact with a computer. Dynamic Braille displays perform a similar function for situations where more discreet communication is required. However, these technologies generally only allow access in a linear manner (from the top left corner of the screen). Further to this, non-textual information such as pictures and diagrams are not easily displayed in this manner. The goal of the work described here is to examine techniques to enable users to explore information or shapes non-visually and to navigate computer interfaces in a non-linear manner.

One potential solution to these problems investigated here is to use tactile cueing to allow a visually impaired user to browse data. Coded tactile representations such as Braille have been used successfully for many years to transfer text information into a non-visual form. Tactile cues offer the potential of a more general form of structured tactile message to present information through the tactile channel. To provide useful and usable tactile messages, it is important that the user can quickly and easily distinguish between each message to extract the contents. Here we describe a series of studies that examine performance in distinguishing pin array tactile cues encoding one or many pieces of information. At this stage of the work, we are particularly concerned with how to present the information in a useful manner. In particular, we will extend previous work on Tactons [3] – structured abstract tactile messages that encode information in the different parameters of the tactile signal – from a vibrotactile representation to a pin array representation. This paper describes a series of experiments that have been conducted to examine the appropriate design of pin array Tactons. For the purposes of the remainder of this paper, the word Tactons will be used to refer to pin array Tactons.

Here we discuss these techniques from the perspective of providing interfaces for visually impaired computer users. Similar problems exist in situations where a sighted user has restricted visual feedback. This could be for example in mobile situations where the user is concentrating on safely navigating a busy environment. The techniques described here could also potentially be of use in

- ‡ are with the Laboratoire d'Informatique Théorique et Appliquée, Université Paul Verlaine — Metz, UFR MIM, Île du Saulcy, 57006 Metz, France {pietrzak, benoit.martin, pecci}@univ-metz.fr
- † are with the Department of Computing Science, University of Glasgow, 18 Lilybank Gardens, Glasgow G12 8QQ, Scotland {ac,Stephen}@dcs.gla.ac.uk

Manuscript received (insert date of submission if desired). Please note that all acknowledgments should be placed at the end of the paper, before the bibliography.

these situations.

In section 2, related work in the area is described, including a definition of Tactons and relevant research from the accessibility field. Section 3 describes how we extend the idea of Tactons to pin array devices with various parameters for displaying information being considered. A series of experiments are presented in section 4 where the performance of sighted users is examined when presented with different forms of Tactons. Section 5 extends this work to include Tactons where multiple dimensions of the signal are varied simultaneously to display information. Evaluations of two applications are described in section 6, with visually impaired people using interfaces that employ Tactons. Finally in section 7, the results of the experiments are discussed as a whole and conclusions are drawn.

## 2 RELATED WORK

Traditional methods of accessing diagrams non-visually use raised paper, which lifts certain parts of the image to allow users to explore shapes or lines presented through tactile relief. Providing accessible tactile diagrams through this method is not a trivial task however. Many authors have noted that a direct translation of a visual diagram to a tactile diagram is in most cases not sufficient to provide accessible tactile diagrams (for example [7], [5], and [10]). While tactile diagrams provide an invaluable tool for allowing visually impaired people to browse non-textual information, they suffer a number of disadvantages. Firstly they are static representations that it is difficult to change without reprinting the image. Secondly they rely solely on tactile relief and cannot take advantage of any computer based technologies such as screen readers or dynamic tactile devices to aid comprehension.

As such, a number of attempts to provide computer based or hybrid alternatives to raised paper have been investigated. Wall and Brewster [18] present a computer based system for accessing bar charts that shares many features with a raised paper diagram. The user navigates the image by moving a stylus over a graphics tablet representing the physical piece of raised paper. The user's non-dominant hand rests on a raised pin tactile display that provides a simple binary up or down signal to the user for the area around the user's cursor depending on whether they are above a bar on the graph or over free space. One immediate advantage of this system over a traditional raised paper representation is that it is computer-based. Charts can easily and quickly be reloaded. The system can take advantage of the computer-based representation to track the user's movements and provide further feedback to aid the user to navigate the environment.

There are several commercially available dynamic pin array devices available. Bliss et al. [1] describe the Optacon, an early device that was designed to make printed information accessible to visually impaired people by combining a camera and vibrotactile array. The user moved the camera over a document with his or her dominant hand. The printed information was then displayed to the user's non-dominant hand through a vibrotactile pin array with the dark areas of the document represented as vibrating pins and the light areas as stationary pins. The device used in the studies described here is the VTPlayer tactile mouse (shown in Figure 1), which has two 4x4 arrays of blunt pins designed to rest under the users' index and middle fingers as they hold the mouse. The pins have a diameter of 1mm, and the space between two pins is 1mm. The pin arrays on this device are sized for a user's fingertips, which is small compared to some other commercial devices such as the Dot View tactile display developed by KGS Corporation (www.kgs-jpn.co.jp). One advantage of using the VTPlayer over a larger device is that the cost of a small pin array is considerably lower than a larger array. Using a small array also ensures that users can feel the tactile signals with one hand, and can cover a whole array with a finger meaning that they will not miss tactile signals when touching the array.

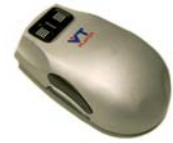

*Figure 1. The VTPlayer tactile mouse*

By far the most widely used form of tactile cueing system is Braille, which uses six raised dot positions arranged in a 3x2 rectangle to codify characters. Developed in the 19th century, it has been used successfully for many years to allow blind and visually impaired people to read text. The most common medium for Braille is a static raised paper representation, although dynamic Braille devices can now be used to present messages though a dynamically changing array of pins. It is excellent for any information that can be represented easily as text, but is less useful for non-textual information.

In this paper, we attempt to formalize the design of a more general form of pin array tactile messages. To achieve this, we combine a pin array style coded representation with work on vibrotactile Tactons. Brown et al. [3] define Tactons as "structured tactile messages" and are analogous to Earcons in audio [2]. Previous vibrotactile Tacton studies from Brown et al. [3] have examined vibrotactile devices as a method of providing tactile cues to users. They initially examine one-dimensional vibrotactile Tactons and show how a high degree of accuracy can very quickly be reached for simple vibrotactile Tactons. They then demonstrate how using multiple independently identifiable parameters of the signal – chosen such that they do not interfere with the recognition of other parameters – can be used to increase the flow of information to the users. Multiple vibrotactile actuators were attached to different locations on a user's forearm. Information was displayed to the user through three parameters of the tactile signal: the rhythm at which the actuator is vibrated, the roughness of the vibration and the body location of the actuator stimulated. Further to this, Brown et al. [4] demonstrate that users could achieve a high level of accuracy with three varying parameters of the tactile signal. However, they note that a careful choice of pa-rameters is required to ensure that the parameters chosen do not interact when varying their values. For example, they note that increasing the frequency of a tactile signal can also lead to a perceived increase in amplitude in the signal. Further work by Hoggan and Brewster [8] has examined the cross-modal equivalence between vibrotactile Tactons and auditory Earcons. This would allow the same information presented in different modalities in different situations.

Many other examples of structured vibrotactile messages exist particularly in the mobile device field, including Vibetonz from Immersion corp. Using the Vibetonz system, more information can be given to the user through the vibration than just a simple buzz to alert them of an event. Similarly Chang and O'Sullivan [6] have studied the addition of vibrotactile feedback into a mobile device in order to enhance the audio feedback. Further to this, they attempt to provide a classification of haptic sensations within a generalized framework [14]. One closely related study is described by Megard et al. [12] who examine users' spontaneous associations between tactile patterns formed with a vibrotactile array (the VITAL display developed by CEA LIST) to verbal descriptions for the purposes of building tactile map legends. Tactile cueing is also frequently used to present information to users in situations when a person's vision is otherwise occupied. Van Veen and van Erp describe the use of a tactile bodysuit to present orientation information to a fighter pilot in-flight [17]. A line of tactile actuators is activated over the pilot's torso to indicate their relative orientation to the horizon. Related work into the parameters of tactile perception is discussed by MacLean and Enriquez [11]. They describe a series of experiments examining usable parameters for haptic icons or 'Hapticons'. To display the Hapticons, their studies used a one degree of freedom force feedback wheel which the user grasped between the thumb and a finger. The parameters examined to display data included frequency of vibration, waveform shape and force magnitude.

## 3 DESIGN OF TACTONS

There are two ways to display information on a tactile pin array. The first one is to directly convert pixel shade to pin position: light pixels are converted into raised pins, and dark pixels into lowered pins. This method has been used by Jansson and Pedersen to present blind users a tactile map with a VTPlayer mouse [9]. They concluded that the tactile information did not provide significant performance advantages to an audio-only environment.

We present a second solution that uses a coded representation to present information with a pin array. We use tactile icons that we call pin-array Tactons. The goal is to suggest information with a pattern or an animation created with a pin array.

### 3.1 Definitions

Here we provide definitions that are used to describe the Tactons in this paper. Given an $n \times m$ size pin array, each pin could be controlled individually and has two states: up and down.

The *pattern* is the state of each pin of the matrix at a given time. Each pin could either be up or down.

A *static Tacton* is only defined by a pattern. It is displayed until the Tacton changes.

A *frame* is a step of an animation. It consists of a pattern, and a *duration*. The duration is unitless: it is only used to compare the durations of different frames.

A *dynamic Tacton* t is an animation. It is composed of a list of frames, and a *tempo*. The tempo is used in addition to the frames' durations to compute the time each frame is actually displayed.

### 3.2 Tacton parameters and dimensions

The *parameters* refer to the Tactons' physical properties defined in the previous section: pattern, frame, frame duration, tempo, and frame list. For example, the parameter pattern can be described by the position (in mm) and state (up or down) of the array of pins such that it could be reproduced exactly by its description. Similarly for tempo, we can specify an exact value in milliseconds.

The *dimensions* of a Tacton set correspond to logical properties based on the parameters. Examples of dimensions are: shape, size, blink speed, animation speed, etc. They are finite sets, so when we define a dimension, we have to define its values too. For example *size = {big, small}* is a dimension, but is not reproducible without reference to one or more physical properties.

The *information* is the message that we want to transmit to the user with Tactons.

The *Tacton sets* are built by combining its dimensions. Other dimensions could be found in the sets, but may not be taken into account as they do not engender Tactons (for example size in set *4* in Figure 2).

To convey information to the user, we structure a Tacton using one dimension for each piece of information. For example, given an object, we want to represent its shape, color and size. There are three pieces of information, so three dimensions are used to represent it. The Tacton's shape could be used to represent the object's shape, the Tacton's size to represent the object's size and the Tacton's blinking speed to represent the color. Each value of the dimension matches a value in the information. So if the object could have two shapes, we must define two shapes in our 'shape' dimension and similarly for the size and the color.

## 4 ONE-DIMENSIONAL TACTONS

The first step when investigating the usefulness of this interaction technique is to evaluate the users' ability to distinguish simple kinds of Tactons. Thus we decided to begin with one-dimensional Tactons. Moreover the goal of these experiments is to find parameters that can be used to build more elaborate Tactons. Here we present a continuation of the work described in [15]. One potentially useful area for supporting non-visual browsing of spatial data is to use these Tactons for direction information, so initially static and dynamic Tactons representing 8 directions were designed. These were: North, South,

East, West (called radial Tactons), and towards the four corners (called diagonal Tactons). There is only one piece of information to be represented so only one dimension is needed. The Tactons designed can be classified into three categories. The first is the static Tactons, and the two others are dynamic Tactons: blinking and waves. The blinking Tactons are dynamic Tactons that alternate a pattern and a frame with no pin up, and the waves are animations that represent the evolution of a shape in space and time. In these studies, the tempo of every dynamic Tacton of every set is 100ms.

### 4.1 Methodology

The following sections describe a series of four studies to identify Tacton sets that are distinguishable. A similar methodology was used in each case so we describe the common elements of the studies first. The Tactons were displayed on the forefinger of the participants' dominant hand. The users' hand and the VTPlayer were hidden inside a box to prevent any visual cues that could affect the results. The task set for the user was to identify the direction indicated by the Tacton presented from the set of 8 possible directions. Each block of tests began with a training session, where the user could explore the 8 Tactons of the set being tested. When they felt that they knew the Tactons well (typically less than 2 minutes), they pressed a key to begin the actual test. In each block of tests, each user was presented with 100 random Tactons and had to identify the pattern. The Tactons were presented on the pin array of the VTPlayer mouse, using both static Tactons and dynamic Tactons. When the users identified the pattern, they moved in the direction that the pattern represented and clicked in the zone corresponding to this direction on the screen. They then moved back to the center of the screen. Then the next Tacton was presented once this was completed.

In each experiment, all users tested all Tacton sets used in the experiment in separate blocks, in a counterbalanced order. Both the number of errors in identifying a Tacton and the average time to answer were measured. After the experiment, participants filled in a questionnaire to provide information on their preferences and to provide suggested improvements. They were not made aware of their performance in the task until after the experiment was completed.

Due to the discrete and bounded nature of the data, non-parametric analysis of the data was carried out. Kruskal-Wallis rank sum tests were therefore used to find significant differences between samples. Pairwise comparisons using Wilcoxon rank sum test, with a Holm p-value adjustment method have been used to identify significant differences between pairs.

### 4.2 Experiment 1 – Comparing Static, Blinking and Wave Tactons

The first study examines the merits of different styles of Tacton: static, blinking and waves. The initial sets were developed iteratively through informal pilot testing, using simple lines and line combinations to generate the patterns. Patterns with different numbers of pins were examined to test whether a larger number of pins improved recognition. Nine users between 24 and 48 years old took part in the first experiment to test Tacton identification of 4 Tacton sets. All participants were sighted with no tactile impairments, and recruited from the department of Computer Science at the University Paul Verlaine — Metz.

The four Tacton sets were chosen to test different factors that may influence identification of the Tactons. The first two sets use a low number of pins (two pins) for each direction (Figure 2). The difference between them is that set 1 is composed of blinking Tactons, whereas set 2 is composed of static Tactons. Set 3 was chosen as a wave-like dynamic set with the same number of frames for each direction. Set 4 is a static set, with far more pins making up the Tacton than the other static set. In each set we tried to maintain a consistency between each Tacton. So sets 1 and 2 used the same number of raised pins for each Tacton, set 3 used the same number of frames, and set 4 used as far as possible the same number of pins.

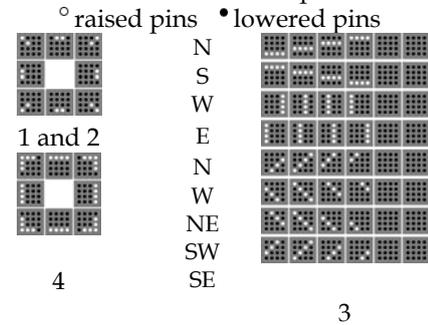

Figure 2. The four sets of Tactons used in experiment 1. Set 1 is to evaluate blinking Tactons, sets 2 and 4 static Tactons with different numbers of pins, and set 3 wave Tactons.

The goal of this experiment is to test the initial design of the Tactons. In particular, this study evaluates whether static Tactons with few pins are harder to recognize than those with more pins, as well as whether the static or the dynamic Tactons are easier to distinguish. Finally, the study evaluates user performance with blinking Tactons compared to static or wave Tactons.

#### 4.2.1 Results

Figure 3 summarizes the results of the experiment. The results show that participants were more accurate identifying the Tactons from set 4 with a median error rate of 0%, compared to 17% for set 1, 8% for set 2 and 3% for set 3. The Kruskal-Wallis test shows a significant difference between the four sets ($\chi^2 = 19.93$, $p < 0.001$). The pairwise tests reveal that set 4 leads to significantly fewer errors than the other sets ($p < 0.002$ for sets 1 and 2, and $p = 0.04$ for set 3). No other significant differences were present in the data. Moreover no significant differences were detected between directions ($\chi^2 = 7.33$, $p = 0.39$; $\chi^2 = 7.64$, $p = 0.26$; $\chi^2 = 5.53$, $p = 0.59$). Set 4 was recognized significantly more quickly than the other sets with a median answer time of 1.66s, compared to 2.64s for set 1, 2.69s for set 2 and 2.74s for set 3. A significant difference can be seen between the sets using a Kruskal-Wallis test ($\chi^2 = 16.32$, $p < 0.001$). Once more the pairwise tests only show a significant difference between set 4 and the others ($p = 0.002$ for set 1, and $p = 0.01$ for sets 2 and 3).

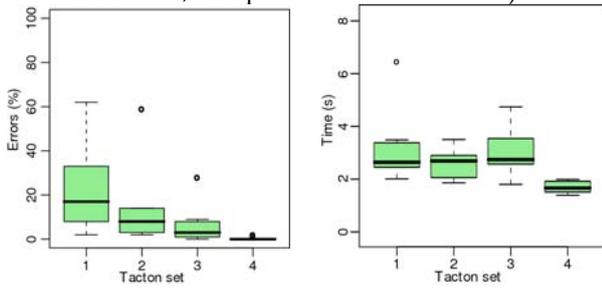

Figure 3. Results of experiment 1

### 4.2.2 Discussion

Sets 3 and 4 use more pins than sets 1 and 2. However as there is no significant difference between set 3 and sets 1 and 2, we cannot assert the Tactons with more pins are better. In the same way, set 2 is a static set, but did not have significantly better results than set 3, so we cannot say that the static Tactons are easier to recognize than the dynamic ones. However, set 4 is easier to use than the other sets in this experiment: users made fewer errors and needed less time to explore them. Blinking Tactons led to a high error rate. This may be due to the fact that the pattern chosen used few pins, as the static Tactons using the same patterns also led to a high error rate. More easily distinguishable patterns need to be tested before discarding blinking Tactons. Moreover, further work is required to investigate the blink speed and the rhythm between the pins up time and pins down time to determine an appropriate range for the parameter (see section 5). This experiment has demonstrated a set of usable static Tactons. These results were therefore used in the following experiment to try and improve on the design of the Tacton sets.

### 4.3 Experiment 2 – More Pins and Growing Shapes

The users of the previous experiment were asked to propose new sets of Tactons, according to what they liked or wished after the tests. After new pilot studies, two new static Tacton sets were selected (6 and 7 in Figure 4), and one new dynamic set (5 in Figure 4). Experiment 2 aimed to compare these three new sets with the best one of the previous experiment, (set 4 in Figure 2). Eleven users recruited among the master students at the University Paul Verlaine — Metz, aged 23 to 27, took part of the second experiment. None had participated in the previous experiment. We have selected the static sets using "a lot of pins", as the best sets from the previous study (sets 3 and 4) use "a lot of pins". Users also suggested they prefer Tactons with more pins. The goal is to find Tactons that are easily and quickly recognizable, as well as pleasant for the user. The dynamic set selected used a new technique - a growing moving shape - as this could potentially help differentiate the directions.

This experiment evaluates new Tacton sets as well as the best one from the previous experiment in order to find a more efficient wave Tacton set, and to find guidelines for Tacton design. The hypothesis is that Tactons with more pins are easier to identify than Tactons with fewer pins. Therefore better recognition rates are expected from set 6 when compared to recognition rates from sets 4 and 7. We also make the hypothesis that the growing shape of set 5's waves will improve recognition, and then that these Tactons will get low error rates.

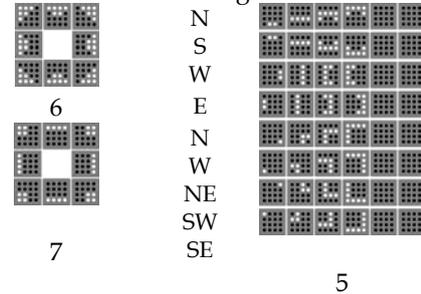

Figure 4. New Tactons used in experiment 2

### 4.3.1 Results

The error rates and the response times of the experiment 2 are summarized in Figure 5. Set 4 was again most accurately identified, with a median error rate of 2%. Set 5 was mis-recognized 23%, set 6 had a median error rate of 14% and set 7 had a median error rate of 10%. The Kruskal-Wallis analysis revealed a significant difference between some of the Tacton sets ($\chi^2 = 10.22$, $p = 0.01$). The pairwise analysis only shows a significant difference between sets 4 and 5 ($p = 0.04$). No significant difference was detected between directions ($\chi^2 = 6.58$, $p = 0.47$; $\chi^2 = 6.80$, $p = 0.44$; $\chi^2 = 2.42$, $p = 0.93$; $\chi^2 = 12.59$, $p = 0.08$). Set 7 was the quickest to be recognized with only a median of 1.91s. Set 4 took 2.33s, set 5 needed 3.91s, and users explored set 6 for a median of 2.49s. The Kruskal-Wallis test shows a significant difference between the four Tacton sets ($\chi^2 = 21.39$, $p < 0.001$). Set 7 was recognized significantly faster than set 4 ($p = 0.03$), set 5 ($p < 0.001$) and set 6 ($p = 0.03$). Set 5 was significantly slower than sets 4 ($p = 0.03$) and 6 ($p = 0.01$) as well as set 7 previously discussed.

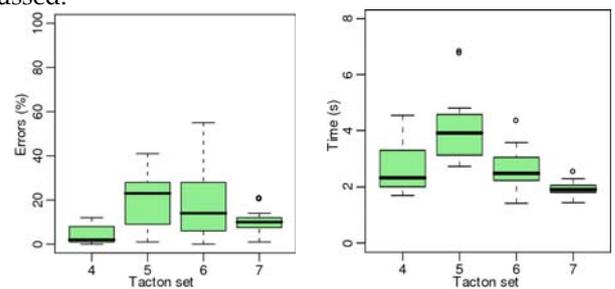

Figure 5. Results of experiment 2.

### 4.3.2 Discussion

The results from set 4 are still positive: they still show a low error rate, and appeared to be significantly better in terms of error rate and exploration time than set 5. In addition to that, set 7 appeared to be significantly faster to recognize than the other sets, including set 4. Tacton set 5 is clearly difficult to discriminate. It confused the users as the shape changes, and the user had difficulty in identifying the steps of the animation. The dynamic sets potentially require more time to recognize since if the information is not understood after one wave, the user must wait to feel another one. Set 5 uses 6 frames of 100ms so each wave requires 600ms, meaning the users needed around 7 waves in average to recognize these Tactons. The recognition time for these Tactons may change depending on the speed parameter of the Tacton. The second hypothesis, asserting that the growing shape will improve the recognition, cannot be accepted, and thus a further study is needed to attempt to identify usable dynamic Tactons (experiment 4). Set 6 had a high error rate as well as a long response time. This could be due to the similarity of the patterns of all the Tactons. Moreover, the location of the shape is difficult to identify as there is no reference point. Sets 4 and 7 use the outer pins and so the shapes are far apart on the display compared to the shapes in set 6. So the high recognition rates of sets 4 and 7 compared to set 6 forces us to reject the first hypothesis that states that Tactons with more pins are easier to recognize than the Tactons with fewer pins. One of the goals of this experiment was to find a more efficient wave Tacton set. This goal has not been achieved, so the goal of the next study is to investigate this point further.

### 4.4 Experiment 3 – Improving Wave Tactons

According to the results of the previous studies, the wave Tactons seemed to be more difficult to identify than the static Tactons. Moreover, some users reported that they had more problems identifying the diagonal Tactons than the radial ones. So to attempt to improve performance using these Tactons, two new dynamic sets were designed, using new diagonals. To be able to compare the results with those of the experiment 1, the same users as in the previous experiment tested these Tactons, in similar conditions.

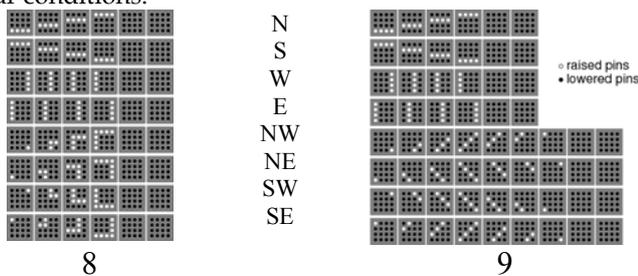

*Figure 6. Tactons used in experiment 3.*

Figure 6 shows the new Tacton sets used for this experiment. Set 8 is a mix of set 3 and set 5: it uses the radials of set 3, and the diagonals of the set 5. The idea is to have a different feeling between radial and diagonal Tactons. Set 9 is modified from set 3. Contrary to set 3, the whole pin array is used to move the line diagonally, using more frames. This means that the animation of the moving diagonal lines takes more frames and longer to complete than the radial lines. The goal of this experiment is to find efficient wave Tacton sets. We hypothesize that the differences introduced between radial and diagonal Tactons will help to discriminate them. So we make the hypothesis that the error rate and that the response time will both be lower than those of previous wave Tactons.

#### 4.4.1 Results

Figure 7 represents the error rates and response times for experiment 4. The results from set 3 of experiment 1 are also compared as set 3 is the best dynamic set identified so far. Results show a median error rate of 0% both for Tacton sets 8 and 9, compared to 3% for set 3 in experiment 1. However the Kruskal-Wallis analysis does not show a significant difference between the sets ($\chi^2$ = 4.67, p = 0.09). No difference was detected between directions ($\chi^2$ = 2.08, p = 0.71; $\chi^2$ = 1.85, p = 0.76). Concerning the time to answer, participants took a median time of 2.74s for set 3 in experiment 1. In this experiment, users needed 2.52s for set 8 and 2.19s for set 9. Participants required a mean of 5 waves to identify a Tacton in set 3, 4 waves for set 8, and approximately 3 for the set 9 (knowing that some Tactons of set 9 have 6 frames and other have 9 frames). However the Kruskal-Wallis analysis does not show a significant difference between the sets ($\chi^2$ = 4.98, p = 0.08).

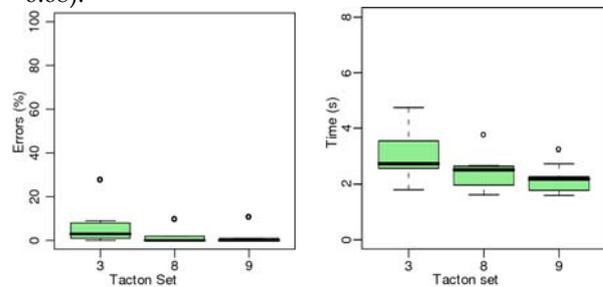

*Figure 7. Results of experiment 3*

#### 4.4.2 Discussion

The majority of the participants reported feeling comfortable with the new wave Tacton sets. Moreover the error rates obtained are low. We can therefore assert that we have found two potentially usable wave Tacton sets. However the difference with set 3 is not significant, the hypothesis about the greater difference between radial and diagonal Tactons allowing easier distinction between them cannot be asserted. Some participants reported feel uncomfortable with this kind of Tactons. One user could not bear the sensation, and stated that he found it extremely difficult to determine the sense of the waves' movement. A future study will investigate how to improve the sensation of these Tactons for this user, for example by slowing down the animation. The next study investigates methods to improve recognition of blinking Tactons.

### 4.5 Experiment 4 – Mixed Tactons

According to users, one of the main recognition problems from the previous study is the lack of a reference point. Indeed, direction is given by the location of the pattern's shape on the pin-array. However, guessing the location is not always obvious. Moreover, blinking Tactons were only used previously in experiment 1. Since good static and wave Tactons have been developed, this study now investigates blinking Tactons in more detail. The same users that took part in experiment 1 evaluated four new Tacton sets in this experiment.

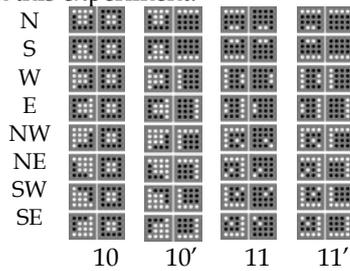

*Figure 8. Tactons used in experiment 4.*

The Tactons used in this experiment are mixed in the sense of one part of the Tacton is blinking, and the other part is static. The direction is given by the same patterns as in set 4 used in the previous experiments, except the diagonals use the whole length of the matrix instead of only 3 pins. Two kinds of reference point have been designed. The first one is a 2x2-pin square placed at the middle of the matrix (sets 10 and 10′, Figure 8). The other is the opposite direction, pointed by a smaller pattern (the one used in sets 1 and 2). It is used in sets 11 and 11′ (Figure 8). For these two kinds of reference points, two techniques have been used. Either the reference point was static and the direction was blinking (sets 10 and 11), or the direction was static and the reference point was blinking (sets 10′ and 11′). The goal of this experiment is to improve the recognition of blinking Tactons. We make the hypothesis that the Tacton sets with a blinking reference point will be significantly easier and faster to recognize than the Tactons with the blinking direction. Moreover we make the hypothesis that the sets which use the directional pattern as a reference point will be significantly easier and faster to recognize than the Tactons that use the middle square as a reference point.

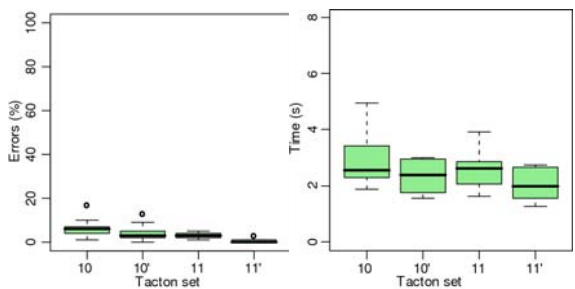

*Figure 9. Results of experiment 4.*

### 4.5.1 Results

The error rates and the average answer times are represented on the charts of Figure9. Users made a median of 6% mistakes for set 10, compared to 3% for set 10′, 3% for set 11 and 0% for set 11′. The Kruskal-Wallis analysis reveals a significant difference between the sets ($\chi 2 = 15.53$, $p = 0.001$). If we look closer with a pairwise test, set 11′ appears to lead to significantly fewer errors than the other sets ($p = 0.006$ for set 10, $p = 0.02$ for set 10′ and $p = 0.01$ for set 11). No other difference is noticeable for errors, and the Kruskal-Wallis test does not show any significant difference for answer time ($\chi 2 = 3.58$, $p = 0.3$). Users spent a mean of 2.56s to recognize the set 10, 2.39s for set 10′, 2.63s for set 11 and 1.99s for set 11′.

### 4.5.2 Discussion

Set 11′ appears to be recognized more easily than the other sets. Participants also expressed a preference for this set in post hoc discussions. Five users out of 9 preferred set 11′, compared to two users for set 11, two for set 10′ and none for set 10. As well as being the least preferred set for the participants in this study, set 10 achieved the poorest recognition rates. Seven of the nine participants preferred Tactons with a blinking reference point rather than the ones with a blinking direction. However the statistics cannot allow us to assert than Tactons with a blinking reference point are better than Tactons with a blinking direction, especially because of the poor results of set 10′. The first hypothesis is therefore rejected. Moreover as the results from set 11 are not significantly different to those from sets 10 and 10′, we cannot conlude that Tactons with a directional reference are better than Tactons with a central reference. So although seven of the nine participants expressed a preference for the directional reference, we reject this second hypothesis. However these Tactons obtained a high level of recognition. Thus we have reached our goal of identifying usable blinking Tactons.

## 4.6 Conclusion

These experiments lead to the development of several sets of distinguishable Tacton sets, using the different techniques proposed: static Tactons, wave Tactons and blinking Tactons. The goal of the first experiment was to compare Tacton sets made with these three techniques. It appeared that users preferred static Tactons, and were able to achieve high recognition rates with static Tactons with many pins raised. We obtained intermediate, but encouraging results from wave Tacton sets. However one user expressed a strong dislike of these Tactons, finding the sensations disturbing. Other wave Tacton sets were tested in the following experiment in order to improve the results. The second experiment evaluated Tactons with more pins, which appeared to be a failure. The shapes were too similar or too close on the matrix. Moreover, wave Tactons were introduced whose pattern changed over time. This was also a failure, since the users had difficulties interpreting the changing shape. The best performing Tacton set from the previous experiment confirmed its good results, and another static set similar to the previous one also obtained good results. A third experiment was also conducted to find better wave Tactons. The results suggest that a usable wave Tacton set was discovered, even if the statistics do not allow us to distinguish them from the previous wave Tacton sets. The user who was not comfortable with the previous wave Tacton sets still was disturbed with these ones. Efforts will be made in future studies to improve the sensation of this kind of Tacton. Finally, the fourth experiment's goal was to try to evaluate different sets of blinking Tactons. A ref-erence point had been introduced in order to help the user to resolve ambiguities. Two modes have been tested: either the direction was blinking and the reference point was fixed, or the inverse. This time blinking sets with low error rates as well as low response time were demonstrated. Now good Tacton sets using each of the techniques have been determined, we now investigate Tactons with several dimensions.

## 5 MULTI-DIMENSIONAL TACTONS

The series of work described in the previous section examined varying one dimension of a pin array Tacton to display information to a visually impaired person or to a person in a situation where vision is restricted. In this section we will examine methods to get more information across to a user through varying more than one dimension of the tactile signal simultaneously. As well as varying the shape of the pattern, we will now vary the tempo of a blinking Tacton and the number of pins forming the pattern to display more information to the user per Tacton than was displayed in the previous study. While increasing the amount of information in the signal has the potential to provide the user with more information, it will also increase the complexity of the signal, and therefore potentially the error rate in recognizing the Tacton. This experiment will examine whether the benefit of increasing the amount of information in the signal outweighs the cost.

To measure the performance of users when identifying the Tactons, we will use the concept of information transmission. Miller [13] describes how information transmission can be used to describe the quality of a communication channel by examining the number of potential messages that can be input into the channel, and the correlation of the input with the user responses. The advantage of using information transmission as a metric is that it provides a unit free method of comparing communications channels. The larger the number of different messages that a channel allows increases the maximum amount of information that can be passed through that channel. However, as the number of messages increases, the complexity of the messages increases which may lead to higher error rates when identifying each message. The higher the error rate, the lower information we say is transmitted through the channel.

### 5.1 Developing the Tactons

When many dimensions of information are present in the signal, it is important that they not interfere such that the user finds it difficult to perceive the individual dimensions.

There are many potential parameters available for use for pin array Tactons. For example, we may consider using the pattern shape, the pattern size, pattern movement, the tempo for a blinking Tacton, or introduce different rhythms by varying the frame durations. However, not all of these will provide benefit. Our previous work described in section 4 has shown, for example, that some dynamic shifting patterns and patterns with less pins can be more difficult to distinguish than static patterns with a lot of pins. For these experiments, we therefore focused on the easier to distinguish patterns. The three dimensions chosen for this initial study were direction, size and speed, using the pattern shape, pattern size and dynamic Tacton tempo parameters. Shape has previously proved to be a successful method of transferring information through pin array devices. Similar shapes were chosen as one dimension of the Tactons used in this study. To test the suitability of size and tempo as useful parameters for transferring information to the user, pilot studies were first conducted individually.

### 5.1.1 Direction

As in the previous studies, eight patterns derived from previous experiment described above were chosen to represent the eight directions. The Tacton sets we use in this experiment use the same shapes as Tacton set 4 (Figure 2): lines for radial Tactons and angles for diagonal Tactons. The directions used for both large and small patterns are shown in Figure 10. Using similar patterns for the large Tactons as in the previous one-dimensional study allows basic comparisons to be made between one and multi-dimensional Tactons. This allows us to gain insight into how the extra complexity brought on by the additional information being presented to the user affects user performance. The direction dimension in this study is therefore defined by *Direction = {N, S, E, W, NE, NW, SE, SW}*, and the parameter used to encode it is the pattern's shape.

### 5.1.2 Size

Given the shapes chosen from the previous study and the limited number of pins available in the VTPlayer mouse array (4x4), only two sizes of shape were possible. The larger ones use the radial Tactons of set 4, and enlarged diagonals from set 4 (Figure 2). The smaller ones use the diagonals of set 4, and shorten radials from set 4. See Figure 10 for the actual patterns used. A short pilot study with four participants showed that four participants were able to distinguish between large and small patterns with minimal training over 90% of the time over 96 trials each. These results suggest that size is potentially a useful dimension that should be investigated further. The dimension in this study is therefore defined by *Size = {small, large}*. It is coded with the size parameter that varies the amount of raised pins, giving the patterns shown in Figure 10.

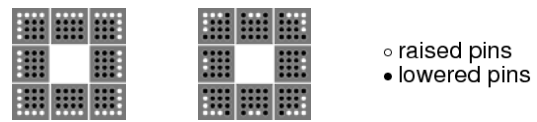

*Figure 10. The large shapes (left) and corresponding small shapes (right) used*

### 5.1.3 Speed

When choosing the speed values, there are a number of different rhythms that were considered. For $d_u$ being the duration that the frame where the pattern is displayed for and $d_d$ being the duration for the frame where no pattern is displayed, we could choose values such that:
- $d_u = d_d$
- $d_u$ varies and $d_d$ is constant
- $d_u$ is constant and $d_d$ varies
- $d_u + d_d$ is constant

In a short pilot study, four participants performed each condition for trials with 7 potential values for tempo from 40ms to 500ms. They were presented with four repetitions of all possible stimulus pairs and in each case had to reply whether the stimuli were the same or different. Only speed was varied and only one stimulus could be experienced at the same time. The results showed extremely similar performance for all four rhythms. $d_u$ being set equal to $d_d$ was chosen as the method for future studies as no preference was suggested by the users and using this method allowed comparisons to be made with the previous study with one-dimensional Tactons.

When selecting what values to use, there will be a trade-off between message length and distance between the values. Participants from the pilot experienced no confusions between 5 of the 6 values chosen. However, as users will be judging the frame duration as a relative measure (*i.e.* fast just means faster than the medium duration as opposed to a specific duration), we will use results from Brown *et al.* [4] that suggest separating three levels of a parameter in their vibrotactile Tactons that vary only relatively to other values already proved difficult. We therefore limit our range of speeds for this study to a maximum of three levels only. The speed dimension in this study is therefore defined as *Speed = {slow, medium, fast}*. The parameter used to represent this dimension is the tempo of the dynamic Tactons. Indeed the rhythm is fixed with 1 as duration for every frame. The tempos for the slow, medium and fast values were 40ms, 200ms, and 500ms respectively

### 5.2 Methodology

A between groups study was run with 20 sighted participants, who were recruited from the student mailing lists at the University of Glasgow. The age range of these participants was 18 to 29. There were four left handed participants split evenly between conditions. Sighted participants are used to provide a baseline performance and to inform the design of future studies with visually impaired participants. In all instances, participants felt the Tactons through the index finger of their non-dominant hand. The non-dominant hand was used as future two-handed focus and context interactions (such as the Tactile Maze and Electrical circuits applications described below) were envisaged. The Tactons were displayed through the VTPlayer tactile mouse with a 4 x 4 array of pins used to display the patterns. The three dimensions varied for each Tacton were '*Shape*', '*Size*' and '*Speed*'. Participants wore headphones playing white noise throughout the study to block any audio cues from the movement of pins on the device, and their hand was hidden to prevent from any visual cues. Headphones were used in this study (unlike in the previously described studies) as this was the first experiment where the tempo was varied and needed to be identified by the participants. Indeed the sound produced by the VTPlayer is due to the movement of the pins, and as the tempo was not varied in the first experiment, the Tactons sounded the same.

There were two conditions tested: a three speed condition ($S_3$) and a two speed condition ($S_2$). Participants were randomly assigned to one condition or the other, with balancing to ensure that an equal number of participants performed each. Due to the fact that $S_3$ has an extra level of speed, there is a larger of range of Tactons available in this condition. Here we choose to maintain an equal number of trials in each condition rather than an equal number of repetitions of each Tacton to avoid potential complications with fatigue and learning effects. There were 32 Tactons in $S_2$ and 48 in $S_3$. Each participant was given 96 Tactons to identify. Participants in $S_2$ and $S_3$ were therefore presented with two or three repetitions of each Tacton respectively.

Participants placed their non-dominant hand on the VTPlayer mouse and held down a key on the keyboard with their dominant hand to feel the Tacton. Once they released the key, the Tacton was stopped. They then gave their answers verbally to the experimenter. The maximum time the Tacton was felt for was capped at 10 seconds. To explore the potential benefits of using multiple parameters in the signal, we will use information transmitted per Tacton as a metric to compare the performance of $S_2$ and $S_3$ with the best Tacton set in the previous study (set 4 in Figure 2). These results from the previous experiment are referred to here with $T_4$. For this study, we hypothesize:

- Participants in $S_2$ will make significantly less errors in identifying Tactons. This due to the fact that participants should find it easier to differentiate two as opposed to three speeds.
- Participants in $S_2$ will take significantly less time to identify Tactons than participants in $S_3$. Again, this is due to the fact that participants should find it easier to differentiate two as opposed to three speeds.
- There will be significantly more information transferred during $S_3$ when compared to $S_2$. Although more errors may be made by participants in $S_3$, there is also the potential for a greater amount of information to be transferred.
- There will be significantly more information transferred to participants in both conditions than in the best case of the one-dimensional Tactons study described in section 4 (set 4 in Figure 2).

### 5.3 Results

The Mann Whitney test was used to test for significant differences in the independent measures between groups. For comparing data within groups for the individual dimensions, Paired Wilcoxon tests were used.

#### 5.3.1 $S_3$ vs. $S_2$

When looking at correct identification of all three dimensions together, there were a median of 12.5% errors in $S_2$ compared to a median of 28.12% in $S_3$. This difference was shown to be significant (W = 135.0, p < 0.03). When analyzing each of the individual dimensions, the differ-ence was shown to be due to the speed dimension. Participants in $S_3$ made a median of 9.9% errors in $S_3$ compared to 0.52% in $S_2$ with this difference being significant (W = 61.5, p < 0.01). No significant differences were found in the Shape (W = 97.5, p = 0.60) or size data (W = 101.5, p = 0.82).

Figure 11 shows the errors made by participants in both conditions for each dimension. There was no significant difference recorded in the time taken to identify the Tactons in $S_3$ and $S_2$ (W = 107.0, p = 0.91). The median time to identify a Tacton for participants in $S_3$ was 2.09s compared with 2.31s in $S_2$.

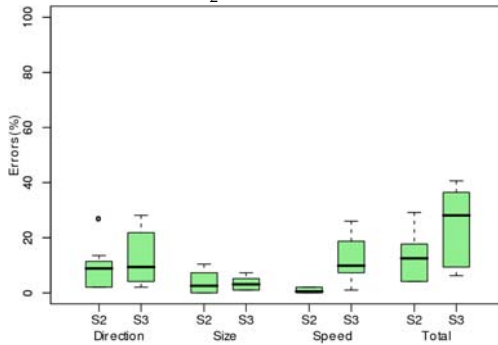

*Figure 11. A box plot showing errors for each dimension and overall, and for both conditions.*

When examining the information transmitted (Figure 12), a median 5.0 bits per Tacton was transmitted in $S_3$, compared with a median of 4.6 bits per Tacton in $S_2$. These data were tested using a Mann Whitney test and this difference was found to be significant (W = 75.0, p < 0.03). This compares with a median information transmission rate of 2.88 bits per Tacton for the best Tacton set tested in the previous study examining pattern only. These data are significantly lower than found in both $S_3$ (W = 153.0, p < 0.01) and $S_2$ (W = 153.0, p < 0.01).

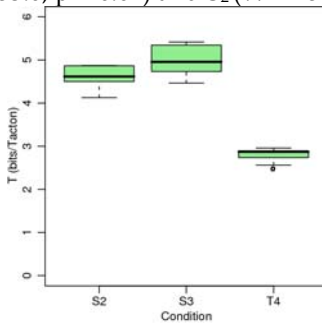

*Figure 12. A box-plot showing the information transmission for all participants in $S_2$, $S_3$, and $T_4$.*

### 5.3.2 Significant Interactions Between Dimensions

Size had a significant effect on participants' performance in both $S_3$ and $S_2$. The large patterns were significantly easier to identify than the smaller patterns. There were significantly more Tactons of small size misrecognized in both $S_3$ (W = 0.0, p < 0.01) and $S_2$ (W = 0.0, p < 0.01). The median percentage of misrecognized small Tactons in $S_3$ was 28.13% compared with a median of 15.63% for large Tactons. This pattern was repeated in $S_2$ with medians of 5.21% and 16.67% misrecognized Tactons for large and small patterns respectively. By analyzing the individual Tacton dimensions, it is shown that this difference is due to significantly more direction errors being present in the data in both $S_3$ (W = 0.0, p < 0.01) and $S_2$ (W = 0.0, p < 0.01). In $S_3$, there were medians of 2.08% and 16.67% errors in distinguishing the directions for large and small patterns respectively. Similarly, there were median errors of 2.08% and 14.58% for large and small patterns respectively in $S_2$. Participants, also required longer feeling the stimuli to answer when identifying small Tactons and large Tactons in $S_3$ (W = 6.0, p < 0.04) and $S_2$ (W = 6.0, p < 0.04). For $S_3$, participants felt the stimuli for medians of 1.94s and 2.27s before answering for large and small patterns respectively. Similarly for $S_2$, participants felt the stimuli for medians of 1.97s and 2.62s before answering for large and small patterns respectively.

### 5.4 Discussion

Firstly, it can be seen that the method used to analyze the data affects how we view the results. Looking simply at error rates, there were significantly fewer errors in the two speed condition when compared to the three speed condition. This is unsurprisingl due to an increase in errors in identifying the speed dimension, but is an interesting result as it enforces the findings of Brown *et al.* [4]. Participants are making relative judgments of a blinking Tacton's tempo and Brown's finding suggested that adding a third level of a parameter that required relative judgments could significantly increase complexity for the user. These data would suggest that it is better to use only two speeds.

However, if we examine the data in this manner, we ignore the benefits of including an extra level of this dimension. It allows us to transfer more information to the user in one Tacton. The benefits are brought out by the Information Transmission results showing that significantly more information per Tacton was transmitted to the users despite these extra errors. Care must still be taken however in interfaces where the cost of the extra errors is high.

The significant interactions shown between size and direction suggest that these dimensions might not be independent on a small tactile pin array, possibly due to the fact that both use the pattern parameter of the signal. Performance was shown to be significantly worse for the small Tactons indicating difficulty identifying patterns with fewer pins. When choosing dimensions for multi-dimensional Tactons, they should not interfere. However, decreasing the size affected how participants perceived the shape of the pattern in both $S_2$ and $S_3$.

## 6 APPLICATIONS USING TACTONS

The previous studies have examined the performance of sighted users in distinguishing a range of different forms of pin array Tacton. Two applications using these techniques are now discussed to illustrate how the concept could be used in an accessible interface. The first is a tactile maze game that has been designed to allow a visually impaired user to navigate a maze using touch alone. The second is an environment designed to teach visually impaired children about electrical circuits through touch.

### 6.1 The Tactile Maze

A maze application was designed to test the Tactons developed in an application setting. Users navigated through a two dimensional maze set in the vertical plane using a PHANToM OMNI device (a three degree of freedom force feedback device from SensAble Technologies) held in their dominant hand, which was used to restrict them to the maze path. No visual feedback was given to the user; however directions to the exit of the maze were presented to the index finger of their non-dominant hand through Tactons presented on a VTPlayer mouse as shown in Figure 13 (left). This bi-manual technique using the PHANToM to navigate was chosen as opposed to a one handed technique using only the VTPlayer mouse due to early work indicating the difficulty of mouse use for visually impaired users [9]. The goal of the game was to follow the directions indicated by the pin array Tactons and reach the exit. There were four possible tactile messages directing the user either up, down, left or right in the maze.

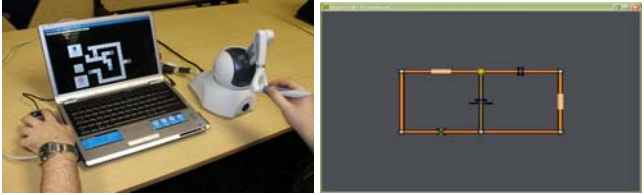

*Figure 13. A user navigating the tactile maze (left) and the circuits application (right).*

An experiment was conducted to compare the use of static and dynamic pin array Tactons for visually impaired users in an application context. The static and dynamic sets of Tactons that were chosen for evaluations are sets 4 and 3 from Figure 2 respectively, although the diagonals were not used as only 4 messages were required. Ten participants with little or no residual vision took part in the study and experienced 10 mazes non-visually in both conditions in a counterbalanced order. Mirror image mazes were used in either condition to ensure the complexity remained the same and valid comparisons could be made. The results showed that participants completed significantly more mazes with the static cues than dynamic cues, and were significantly faster completing the mazes in the static condition. A strong preference was the static cues over dynamic cues by nine of the ten participants. It was felt that the static cues were more similar to what they had previously experienced when using Braille or raised paper. These results agree with the finding of the study with sighted users described above in Section 4. Similarly to the sighted users, visually impaired users performed better with the static Tactons, and expressed a preference for these over the dynamic Tactons. The fact that a high percentage of mazes were completed in both conditions (89% and 72% for static and dynamic conditions respectively) suggests that the participants could successfully interpret the cues in an application setting [7].

### 6.2 Electric Circuits

We developed a multimodal Electric Circuits exploration application, intended for visually impaired people (shown in Figure 13 (right)).

The user can explore a circuit at two levels simultaneously: the global level concerns the recognition of the circuit's shape, and the local level concerns the components recognition. The user is provided with several techniques to aid both exploration levels. Users can navigate in the circuit both with a mouse or a PHANToM OMNI. When navigating with the PHANToM, the user can feel bumps of varying amplitude and direction as force feedback cues or tactile cues using a VTPlayer mouse [16]. As in the maze application, it is bi-manual interaction: the user holds the PHANToM in his or her dominant hand and the VTPlayer in the other. We use the directional Tactons to help the user navigate in the global level if he or she wants help. We construct dynamic Tactons, displaying in sequence the directions available for exploration. The patterns used are from set 4 in Figure 2. At the local level we use other Tactons to encode the components (Figure 14). 6 static Tactons have been designed, using the pattern parameter to encode the information as the experiments in section 4 proved this parameter to be useful. The Tactons represent from left to right a battery, a capacitor, a lamp, a resistor, a junction and a wire. The experiments were conducted with 13 visually impaired children from 9 to 17 years old from several schools around Metz in France. These users explored up to six circuits using the system. The results have shown that users with partial vision prefer to use their remaining sight to complete the task rather than trying to use the tactile feedback. However, blind users had to rely on haptics. They had difficulties understanding the circuit's shape. Few users made errors when recognizing the components with half of the users preferring to use the force feedback cues and the other half preferring the tactile cues.

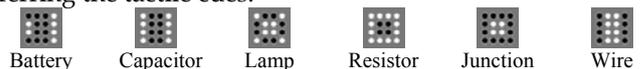
Battery    Capacitor    Lamp    Resistor    Junction    Wire

*Figure 14. Tactons used for the components.*

## 7 CONCLUSIONS

Here, we have followed through the design and evaluation of pin array Tactons. We have introduced definitions that can be used specify different usable parameters more precisely, such that we can build a common language for the different forms of pin array Tacton. In the initial evaluations, different forms of one-dimensional Tacton were examined. We compared static, dynamic wave, blinking, and mixed patterns, with a range of successful patterns being identified. We then extended these result to Tactons containing multiple dimensions of information. Results showed that although users made more errors identifying the Tactons, significantly more information could be transferred to the user when using more than one parameter of the signal. These results provide a baseline for user performance in the task. However, as it is important to extend our studies to our user group we presented two evaluations of interfaces that use these techniques to present information non-visually. The Maze and Electric Circuit interfaces demonstrate how applications can incorporate Tactons to provide a visually impaired user with information about the spatial layout of an environment. From these studies we can produce the following guidelines:

- Using a reference point can help users to differentiate the Tactons
- Radial and diagonal waves Tactons should have a different design to avoid ambiguities
- Designers can increase the amount of information presented through a Tacton by increasing the number of independent parameters in the Tacton.
- Pattern and blink speed are two factors that can be used independently.
- For small pin arrays, patterns with fewer pins are harder to recognize and take longer to distinguish.
- The speeds used (0.04s to 0.5s per frame) did not affect the time it took to recognise the Tacton.

Tools for guiding users around an environment non-visually have much potential for teaching visually impaired school children in a range of subjects. Future work will examine where these techniques could be useful such as in geography to guide a child around a map. Our current work is examining the integration of these cues into a shape recognition environment for teaching visually impaired children geometry with the tactile feedback used to guide the child around the shape, to gain a greater awareness of its spatial layout. The results from this paper suggest there is benefit in further examining pin array Tactons as a method of transferring information to visually impaired computer users. By adopting techniques drawn from vibrotactile Tactons, we have shown how pin array Tactons can be designed to successfully allow a user to interpret the tactile signal. Future work will integrate the pin array Tactons into more general purpose computer environments to allow a user to navigate a computer interface and browse data non-visually.

## ACKNOWLEDGEMENTS

This work is sponsored by the European project MICOLE (IST-2003-511592). We would like to thank the St. Eucaire school in Metz, the Schuman school in Metz, the Santifontaine center in Nancy, and the Royal National College in Hereford for the time they grant us for our experiments.

**Thomas Pietrzak** is a Postdoc at Université Paul Verlaine — Metz, studying information coding using haptic and multimodal interfaces.

**Andrew Crossan** is a research assistant at Glasgow University studying multimodal interfaces, accessibility and mobile interaction.

**Stephen A. Brewster** is a professor and head of the multimodal interaction group at Glasgow University. His work focuses on non-visual interaction applied to both static and mobile situations.

**Benoît Martin** is an assistant professor at Université Paul Verlaine — Metz. His research interests include haptic interfaces, mobile interfaces, text input techniques and hyperbolic visualization.

**Isabelle Pecci** is an assistant professor at Université Paul Verlaine — Metz. Her research interests include haptic interfaces, mobile interfaces and hyperbolic visualization.